\documentclass{aastex63}

\usepackage{dcolumn}
\usepackage{graphicx}
\usepackage{changepage}

\received{\today}
\revised{\today}
\accepted{\today}
\submitjournal{ApJ}

\shorttitle{Cosmic Discordance}
\shortauthors{Di Valentino et al.}

\begin{document}

\title{Investigating Cosmic Discordance}  

\correspondingauthor{Eleonora Di Valentino}
\email{eleonora.divalentino@manchester.ac.uk}

\author{Eleonora Di Valentino}
\affiliation{Jodrell Bank Center for Astrophysics, School of Physics and Astronomy, University of Manchester, Oxford Road, Manchester, M13 9PL, UK}

\author{Alessandro Melchiorri}
\affiliation{Physics Department and INFN, Universit\`a di Roma ``La Sapienza'', Ple Aldo Moro 2, 00185, Rome, Italy} 

\author{Joseph Silk}
\affiliation{Institut d'Astrophysique de Paris (UMR7095: CNRS \& UPMC- Sorbonne Universities), F-75014, Paris, France}
\affiliation{Department of Physics and Astronomy, The Johns Hopkins University Homewood Campus, Baltimore, MD 21218, USA}
\affiliation{BIPAC, Department of Physics, University of Oxford, Keble Road, Oxford OX1 3RH, UK}

\begin{abstract}
We show that a combined analysis of CMB anisotropy power spectra obtained by the Planck satellite and luminosity distance data simultaneously excludes a flat universe and a cosmological constant at $99 \%$ CL. These results hold separately when combining Planck with three different datasets: the two determinations of the Hubble constant from Riess et al. 2019 and Freedman et al. 2020, and the Pantheon catalog of high redshift supernovae type-Ia. We conclude that either LCDM needs to be replaced by a different model, or else there are significant but still undetected systematics. Our result calls for new observations and stimulates the investigation of alternative theoretical models and solutions.
\end{abstract}

\keywords{CMB, cosmological parameters}

\section{Introduction}

Most  current theories of cosmological structure formation are essentially based on three ingredients: an early stage of accelerated expansion (i.e. {\it Inflation}, see \cite{Baumann:2009ds,Martin:2013tda,Lyth:1998xn} for reviews), a clustering matter component to facilitate structure formation (i.e., {\it Dark Matter}, see \cite{Bertone:2004pz}), and an energy component to explain the current stage of accelerated expansion (i.e, {\it Dark Energy}, see \cite{Copeland:2006wr,Peebles:2002gy}). While there is still no direct experimental evidence for this 'cosmic trinity', numerous viable theoretical candidates have been developed. 

Among them, the currently most popular paradigm of structure formation is the Lambda Cold Dark Matter (LCDM) model, recently even acclaimed as the 'standard model' of cosmology (see e.g., \cite{Blandford:2020omc} and references therein).
The LCDM model is based on the choice of three, very specific, solutions: Inflation is given by a single, minimally coupled, slow-rolling scalar field; Dark Matter is a pressureless fluid made of cold, i.e., with low momentum, and collisionless particles; Dark Energy is a cosmological constant term.

It is important to note that these choices are mostly motivated by  {\it computational} simplicity, i.e., the theoretical predictions under LCDM for several observables are, in general, easier to compute and include fewer free parameters than most other solutions. However, computational simplicity does not imply naturalness. Indeed, while the cosmological constant is described by one single parameter (its current energy density), its physical nature could be much more fine-tuned than a scalar field represented by (at least) two parameters (energy density and equation of state). At the same time,  CDM is assumed to be always cold and collisionless during all the many evolutionary phases of the Universe. Some form of interaction or decay must exist for CDM, but this aspect is not considered in the LCDM model, other than  freeze-out from a thermal origin at high temperature. Finally, the primordial spectrum of inflationary perturbations is described by a power-law, therefore parametrized by only two numbers: the amplitude $A_s$ and the spectral index $n_s$ of adiabatic scalar modes. However, since inflation is a dynamical process that, at some point, must end, the scale-dependence of perturbations could be more complicated (see e.g., \cite{Kosowsky:1995aa}). 

For these reasons, the $6$ parameter LCDM model (that, we recall, is not motivated by any fundamental theory)  can be rightly considered, at best, as an approximation to a more realistic scenario that still needs to be fully explored. With the increase in experimental sensitivity, observational evidence for deviations from LCDM is, therefore, {\it expected}.

Despite its status as a conjecture, the LCDM model has been, however, hugely successful in describing most of the cosmological observations. 
Apart from a marginally significant  mismatch with Cosmic Microwave Background (CMB) observations at large angular scales (see e.g., \cite{Copi:2010na}), 
LCDM provided a nearly perfect fit to the measurements made by the WMAP satellite mission, also in combination with complementary observational data such as those coming from Baryon Acoustic Oscillation (BAO) surveys, supernovae type-Ia (SN-Ia), and direct measurements of the Hubble constant (see, e.g. \cite{Komatsu:2010fb}).

More recent data, however, are starting to show some interesting discrepancies with the LCDM model. Under LCDM, the Planck CMB anisotropies seem to prefer a value of the Hubble constant that is significantly smaller than values derived in a more direct way from luminosity distances of supernovae (see e.g. \cite{Riess:2020sih}). At the same time, the combination of the amplitude $\sigma_8$ of  matter density fluctuations on scales of $8$ Mpc $h^{-1}$ and the matter density $\Omega_m$, parametrized by the $S_8 \equiv \sigma_8 \sqrt{\Omega_m/0.3}$ parameter, is significantly smaller in recent cosmic shear surveys than the value derived from Planck data under LCDM (\cite{Asgari:2019fkq}).

Systematics can play a role and the LCDM model still produces a reasonable fit to the data. However, the main ambition of modern cosmology is to identify a cosmological model that can be used as an ideal laboratory  to test fundamental physics and possible 'extensions' thereof. For example, stringent constraints on neutrino physics have been placed using Planck data in combination with BAO and other observables (see e.g. \cite{Aghanim:2018eyx,Palanque-Delabrouille:2019iyz,Ivanov:2019hqk}). The possibility of constraining fundamental physics with such high precision is challenged if the underlying cosmological model does not produce an excellent fit to current data. In practice, current tensions are already presenting a serious limitation to what in recent years has been defined as {\it precision} cosmology.

Recently, it has been shown that these tensions are exacerbated when the possibility of a closed universe is considered \cite{DiValentino:2019qzk,Handley:2019tkm}. Planck CMB angular power spectra, indeed, prefers a closed universe at $99 \%$ CL (see e.g. page 40 of \cite{Aghanim:2018eyx}) and this translates to an even lower Hubble constant and an even larger $S_8$ parameter. Moreover, significant tensions at about three standard deviations now emerge between Planck and BAO data (\cite{DiValentino:2019qzk},\cite{Handley:2019tkm}). In practice, not only tensions with cosmological data exist, but even larger discordances may be hidden by the assumption of the LCDM model itself. 

The main problem for a closed Universe is the lack of concordance with other observables. Apart from BAO, indeed, the closed model preferred by Planck does not agree with luminosity distance measurements of supernovae type Ia and predicts too large a matter density $\Omega_m\sim 0.5$ in striking contrast with local measurements of galaxy clustering (\cite{DiValentino:2019qzk}).
However, as we discussed above, many of the assumptions in LCDM, such as the assumption of a cosmological constant, are questionable and lack any robust justification. Hence it is useful to pose the question whether a further increase in the number of parameters, in addition to curvature, can help reconcile the Planck result with other observations.
The main goal of this {\it Letter} is to answer  this question and, to investigate  whether an alternative cosmological model exists wherein current independent cosmological observables are in better agreement than in the LCDM model. In brief, given current tensions with luminosity data and the CMB preference for a closed universe, we search for a new cosmological concordance model that significantly differs from LCDM.

Our approach is simply based on an extension of the cosmological parameter space as we have done before in \cite{DiValentino:2016hlg,DiValentino:2019dzu}. Instead of the usual six parameters, we also allow variations of the dark energy equation of state, the curvature of the universe, the neutrino mass, and the running of the spectral index of primordial fluctuations. 
Finally, in order to check for the robustness of our conclusions, we also investigate the possibility of a systematic in the Planck angular spectra data and that this systematic could be well described by the $A_{lens}$ parameter (see e.g. \cite{Calabrese:2008rt}) that artificially changes the lensing amplitude in the CMB spectra.

\section{Method}

The LCDM model is based on six free parameters (see e.g. \cite{Aghanim:2018eyx}): the angular size of the sound horizon at decoupling $\theta_{MC}$, the cold dark matter and baryon densities $\Omega_c h^2$ and $\Omega_{b}h^2$, the optical depth at reionization $\tau$, and the amplitude $A_s$ and the spectral index $n_s$ of inflationary scalar perturbations. As discussed in the introduction, several tensions between cosmological observables are starting to emerge when this model is assumed. We, therefore, investigate the following extensions (considering them  all {\it simultaneously}):

\begin{itemize}

\item The {\bf curvature parameter $\Omega_k$}. The possibility of a curved universe is fully compatible with GR and is also allowed in some non-standard, inflationary models. Moreover, as  stressed in \cite{DiValentino:2019qzk},  Planck angular spectral data alone prefer models with positive curvature ($\Omega_k<0$, see page 40 of \cite{Aghanim:2018eyx}).

\item The {\bf running of the spectral} index of inflationary perturbations {\bf $\alpha_s = d n_{s} / d ln k$}. A sizable running is expected in many inflationary models, ranging from $\alpha_s \sim (1-n_S)^2 \sim 10^{-3}$ in slow-roll models (see e.g. \cite{Garcia-Bellido:2014gna}) to higher values (see e.g.\cite{Easther:2006tv,Kohri:2014jma,Chung:2003iu}). An indication at about $\sim 3$ standard deviations for a negative running has been recently claimed by \cite{Palanque-Delabrouille:2019iyz} combining Planck with BAO and Lyman-$\alpha$ forest data.

\item The {\bf dark energy equation of state $w$}. We consider a dark energy equation of state of the form $P=w\rho$, where $P$ and $\rho$ are the dark energy pressure and density and $w$ is a free parameter, constant with redshift. $w=-1$ corresponds to a cosmological constant.

\item {\bf The sum of neutrino masses $\Sigma m_{\nu}$}. We know from oscillation and long-baseline neutrino experiments that neutrinos have to be massive. However, the total mass is still unknown. In the LCDM model,  a minimal mass of $\Sigma m_{nu} =0.06$ eV is assumed. However, the total mass  of neutrinos can be higher.

\item {\bf The lensing amplitude $A_{lens}$}. In order to check the robustness of our results we also consider in some specific runs (see below) the $A_{lens}$ parameter \cite{Calabrese:2008rt} that artificially varies the lensing amplitude in the theoretical CMB angular spectra. While unphysical, this parameter could mimic the presence of a systematic in Planck angular spectra data.

\end{itemize}

Concerning the experimental data, we consider:

\begin{itemize}

\item The Planck $2018$ temperature and polarization CMB angular power spectra. In this paper we use the reference likelihood from the Planck 2018 release that is given by the multiplication of the \texttt{Commander}, \texttt{SimALL}, and \texttt{PlikTT,TE,EE} likelihoods (see page $3$ of \cite{Aghanim:2019ame}). This corresponds to the reference dataset used in the Planck papers. We refer to this data simply as {\bf Planck}.

\item The Baryon Acoustic Oscillation data from the compilation used in \cite{Aghanim:2018eyx}. This consists of data from the 6dFGS~\cite{Beutler:2011hx}, SDSS MGS~\cite{Ross:2014qpa}, and BOSS DR12~\cite{Alam:2016hwk} surveys. We refer to this dataset as {\bf BAO}.

\item The luminosity distance data of $1048$ type Ia supernovae from the PANTHEON catalog \cite{Scolnic:2017caz}. We refer to this dataset as {\bf Pantheon}.

\item The most recent determination of the Hubble constant from Riess et al. 2019. This is assumed as a gaussian prior on the Hubble constant of $H_0=74.03\pm1.42$ km/s/Mpc. We refer to this prior as {\bf R19} \cite{Riess:2019cxk}.

\item The recent determination of the Hubble constant from the Tip-of-the-Red-Giant-Branch approach (Freedmann et al. 2020). This is assumed as a gaussian prior on the Hubble constant of $H_0=69.6\pm2.0$ km/s/Mpc (we sum statistical and systematic errors in quadrature). We refer to this prior as {\bf F20} \cite{Freedman:2020dne}.

\end{itemize}

The comparison between theory and data is made adopting the public available \texttt{CosmoMC} code based \cite{Lewis:2002ah} on a Monte Carlo Markov chain algorithm. The theoretical predictions are made using the \texttt{CAMB} Boltzmann integrator \cite{Lewis:1999bs}.

\section{Results}

\begin{figure}
\centering
\includegraphics[width=.4\textwidth,keepaspectratio]{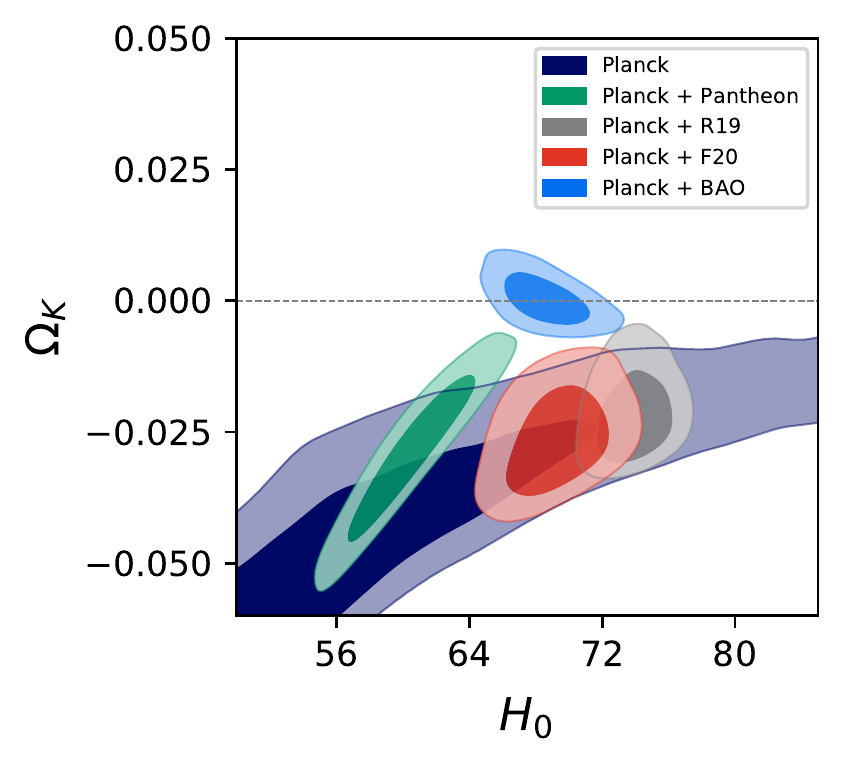}
\caption{{\bf Cosmic Discordance}. Constraints at the $68\%$ and $95\%$ CL on the $\Omega_k$ vs $H_0$ plane for the Planck, Planck+R19, Planck+F20, Planck+BAO, and Planck+Pantheon datasets. 
A $10$ parameters model, LCDM+$w$+$\Omega_k$+$\alpha_S$+$\Sigma m_{\nu}$, is assumed in the analysis.\label{fig1}}
\end{figure}

\begin{figure}
\centering
\includegraphics[width=.4\textwidth,keepaspectratio]{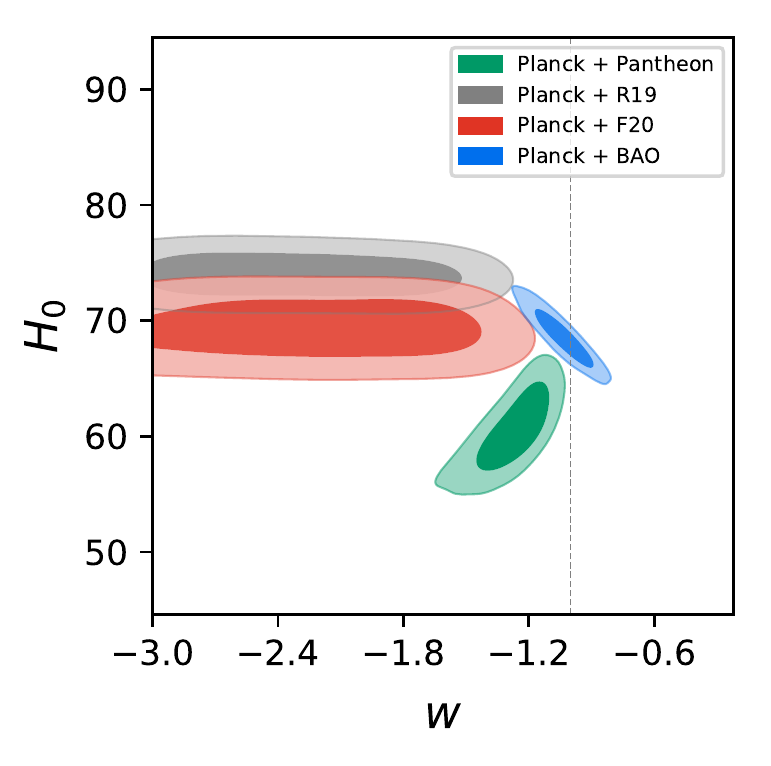}
\caption{{\bf Cosmic Discordance}. Constraints at the $68\%$ and $95\%$ CL on the $w$ vs $H_0$ plane for the Planck+R19, Planck+F20, Planck+BAO, and Planck+Pantheon datasets. A $10$ parameter model, LCDM+$w$+$\Omega_k$+$\alpha_S$+$\Sigma m_{\nu}$, is assumed in the analysis.\label{fig2}}
\end{figure}

Let us first discuss the results when $A_{lens}=1$, as reported in the first five columns of Table~\ref{tab1} and in Figure~\ref{fig1}-\ref{fig2}. For the moment, we therefore neglect the possibility of a systematic in Planck CMB angular spectra. In this case, the preference from the Planck measurements for a closed universe at more than $95 \%$ CL is clearly present in the extended parameter space we are considering.  The confidence levels from Planck plotted in Figure~\ref{fig1}, while very broad and virtually unable to constrain the Hubble constant ( $H_0=53^{+30}_{-20}$ km/s/Mpc at 95\% CL), are clearly below the $\Omega_k=0$ line that describes a flat universe.
On the other hand, the inclusion of the equation of state $w$ now allows the Planck data to be in perfect agreement with the Pantheon, R19, and F20 measurements. As we can see from Figure~\ref{fig1}, all the $95 \%$ confidence regions from the Planck+Pantheon, Planck+F20, and Planck+R19 datasets are well below the $\Omega_k=0$ line. This clearly shows that the recent claims of a closed universe as being incompatible with luminosity distance measurements are a direct consequence of the assumption of a cosmological constant. As we can see from Figure~\ref{fig2}, where we show the 2D contour plots in the $H_0$ vs $w$ plane, all the three luminosity distance datasets, when combined with Planck, exclude a cosmological constant and prefer $w<-1$. In practice, after numerical integration of the marginalized posterior, we have found that Planck+Pantheon, Planck+R19, and Planck+F20 all exclude a cosmological constant and a flat universe at more than $99 \%$ CL.

It is, however, important to highlight that luminosity distance measurements, when combined with Planck, provide values of the Hubble constant that are in tension between themselves. Indeed, it is evident from Figure~\ref{fig1} that the confidence regions from Planck+Pantheon, Planck+F20, and Planck+R19 are inconsistent at more than $95 \%$ CL, providing different constraints on the Hubble constant.

Despite the large parameter space considered, BAO data are in significant tension with the Planck measurements. This can be quickly noticed from the best fit $\chi^2$ value reported in Table I, which increases by $\Delta \chi^2\sim 15$ when the BAO are combined with Planck. The BAO dataset consists of $3$ independent measurements and $8$ (correlated) datapoints. If we consider $5-6$ degrees of freedom, this $\Delta \chi^2$ value suggests tensions at around three standard deviations, in agreement with the findings of \cite{DiValentino:2019qzk,Handley:2019tkm} but now in the case of a LCDM+$\Omega_k$ $10$ parameters model. We therefore stress that the combination of the Planck and BAO dataset should be considered with some caution. Nonetheless, if we force a combined analysis of the two measurements, we can see from Figure~\ref{fig1} that Planck+BAO prefers a flat universe and a Hubble constant compatible with the F20 value. However, Planck+BAO is now not only in tension with Planck, but also with Planck+R19, Planck+F20, and Planck+Pantheon on curvature. Figure~\ref{fig1} clearly reveals the significant tensions present between the current cosmological datasets in our extended cosmological model scenario.

\begin{table}
\footnotesize
\begin{adjustwidth}{-2.8cm}{}
\scalebox{0.74}{
\centering
\begin{tabular}{c|ccccc|cccccccccc}    
\hline\hline                                                                                                          
Parameters & Planck   & Planck + R19 & Planck + F20& Planck + BAO & Planck + Pantheon & Planck + R19 & Planck + BAO & Planck + Pantheon \\ 
\hline

 $\Omega_b h^2$ & $    0.02253 \pm 0.00019$ &  $    0.02253^{+0.00020}_{-0.00016}$ & $    0.02255^{+0.00019}_{-0.00017}$ & $    0.02243\pm 0.00016$ & $    0.02255\pm 0.00018$& $    0.02259\pm 0.00017$& $    0.02262\pm 0.00017$& $    0.02259\pm 0.00017$ \\
 
$\Omega_c h^2$ & $    0.1183\pm 0.0016$  & $    0.1187^{+0.0015}_{-0.0018}$ & $    0.1184\pm 0.0015$ & $    0.1198\pm 0.0014 $& $    0.1186\pm0.0015 $& $    0.1181\pm0.0016 $& $    0.1177\pm0.0015 $& $    0.1181\pm0.0015 $\\

$100\theta_{\rm MC}$ & $    1.04099\pm 0.00035$ &  $    1.04103^{+0.00034}_{-0.00031}$ &  $    1.04105\pm 0.00034$ & $    1.04095\pm 0.00032$ & $    1.04107\pm 0.00034$& $    1.04115^{+0.00032}_{-0.00037}$& $    1.04117\pm0.00033$& $    1.04116\pm0.00033$\\

$\tau$ & $    0.0473\pm 0.0083$ &  $    0.052^{+0.009}_{-0.011}$ &  $    0.0491\pm 0.0079$ & $    0.0563\pm 0.0081$ & $    0.0506\pm0.0082$& $    0.0505\pm0.0080$& $    0.0489^{+0.0089}_{-0.0076}$& $    0.0489^{+0.0085}_{-0.0073}$\\

$\Sigma m_{\nu}$ [eV] & $    0.43^{+0.16}_{-0.37}$ &  $    <0.513$ & $    0.28^{+0.11}_{-0.23}$ & $    <0.194$ & $    <0.420$& $    0.06$& $    0.06$& $    0.06$\\

$w$ & $    -1.6^{+1.0}_{-0.8}$ &  $    -2.11^{+0.35}_{-0.77}$ & $    -2.14\pm0.46$ & $    -1.038^{+0.098}_{-0.088}$ & $    -1.27^{+0.14}_{-0.09}$& $    -1.97^{+0.67}_{-0.59}$& $    -0.88\pm0.10$& $    -1.16^{+0.17}_{-0.11}$\\

$\Omega_k$ & $    -0.074^{+0.058}_{-0.025}$  & $    -0.0192^{+0.0036}_{-0.0099}$ & $    -0.0263^{+0.0060}_{-0.0077}$ & $    0.0003^{+0.0027}_{-0.0037} $& $    -0.029^{+0.011}_{-0.010} $& $    -0.019^{+0.006}_{-0.014}$& $    0.0043^{+0.0038}_{-0.0056}$& $    -0.020\pm0.017$\\

$A_{\rm lens}$ & $    1$  & $    1$ & $    1$ & $    1$& $    1 $& $    1.00^{+0.06}_{-0.13} $& $    1.236^{+0.074}_{-0.090} $& $    1.06^{+0.08}_{-0.13} $\\

${\rm{ln}}(10^{10} A_s)$ & $    3.025\pm 0.018$ &  $    3.037^{+0.016}_{-0.026}$ & $    3.030\pm 0.017$ & $    3.049\pm 0.017$ & $    3.034\pm 0.017$& $    3.032^{+0.018}_{-0.015}$& $    3.028^{+0.019}_{-0.016}$& $    3.029^{+0.018}_{-0.016}$\\

$n_s$ & $    0.9689\pm 0.0054$ & $    0.9686^{+0.0056}_{-0.0050}$ &   $    0.9693\pm 0.0051$ & $    0.9648\pm 0.0048$ &  $    0.9685\pm 0.0051$&  $    0.9705\pm 0.0050$&  $    0.9720\pm 0.0049$&  $    0.9706\pm 0.0050$\\

$\alpha_S$ & $    -0.0005\pm 0.0067$ &  $    -0.0012\pm 0.0066$ & $    -0.0010\pm 0.0068$ & $    -0.0054\pm 0.0068$ & $    -0.0023\pm 0.0065$& $    0$& $    0$& $    0$\\

$H_0 $[km/s/Mpc] & $53^{+6}_{-16}$ &  $   73.8\pm 1.4$ & $   69.3\pm2.0$ & $   68.6^{+1.5}_{-1.8}$ & $   60.5\pm 2.5$& $   73.9\pm 1.4$& $   66.4^{+1.6}_{-1.9}$& $   63.4^{+3.7}_{-5.1}$\\

$\sigma_8$ & $    0.74^{+0.08}_{-0.16}$ &  $    0.932\pm 0.040$ & $    0.900\pm0.039$ & $    0.821\pm 0.027$ & $    0.812^{+0.031}_{-0.018}$& $    0.957^{+0.087}_{-0.043}$& $    0.763\pm0.033$& $    0.819^{+0.023}_{-0.017}$\\

$S_8$ & $    0.989^{+0.095}_{-0.063}$ &  $    0.874\pm 0.032$ & $  0.900^{+0.034}_{-0.031}  $ & $    0.826\pm 0.016$ & $    0.927\pm0.037$& $    0.890^{+0.081}_{-0.041}$& $    0.788^{+0.021}_{-0.018}$& $    0.893^{+0.083}_{-0.074}$\\

$Age {\rm [Gyr]}$ & $    16.10^{+0.92}_{-0.80}$ &  $    14.90^{+0.72}_{-0.32}$ & $  15.22^{+0.054}_{-0.038}  $ & $    13.77\pm 0.10$ & $    14.98\pm0.39$& $    14.81^{+0.95}_{-0.52}$& $    13.67^{+0.13}_{-0.12}$& $    14.56^{+0.74}_{-0.64}$\\

$\Omega_m$ & $    0.61^{+0.21}_{-0.34}$ &  $    0.264^{+0.010}_{-0.013}$ & $  0.300^{+0.017}_{-0.020}  $ & $    0.305\pm 0.016$ & $    0.393^{+0.030}_{-0.036}$& $    0.259\pm0.010$& $    0.321\pm0.017$& $    0.357\pm0.049$\\

\hline

\hline $\Delta N_{data}$  & $0$ & $1$ & $1$ & $5-8$& $1048$& $1$&$5-8$& $1048$\\
$\Delta \chi_{bestfit}^2$ & $    0.0$ &  $    0.62$ & $  0.88  $ & $    14.77$ & $    1037.82$&  $    1.02$&  $    6.53$&  $    1036.87$\\
\hline\hline                                                  
\end{tabular}
}
\end{adjustwidth}
\caption{Constraints at 68$\%$~CL errors on the cosmological parameters in case of the $10$ parameters model LCDM+$w$+$\Omega_k$+$\alpha_S$+$\Sigma m_{\nu}$ on the left, and a $9$ parameters model LCDM+$w$+$\Omega_k$+$A_{lens}$ on the right, using different combinations of the datasets. The quoted upper limits are at 95$\%$~CL. Please note that the Planck alone constraint on the Hubble constant is strongly non Gaussian. In the two bottom lines we quote the increase in the number of datapoints and the best-fit $\chi^2$ values with respect to the Planck dataset alone.}
\label{tab1}  
\end{table} 

From the results in Table~\ref{tab1} we  also note that none of the data combinations considered suggests a negative running. A running of $\alpha_s\sim -0.01$, as claimed in \cite{Palanque-Delabrouille:2019iyz}, is however compatible within two standard deviations with all the datasets. The bounds on the neutrino masses from Planck and luminosity distance measurements are all much more relaxed with  respect to the Planck+BAO case, with the Planck+F20 case even mildly suggesting a neutrino mass of $\Sigma\sim 0.28$ eV.

\begin{figure}
\centering
\includegraphics[width=.4\textwidth,keepaspectratio]{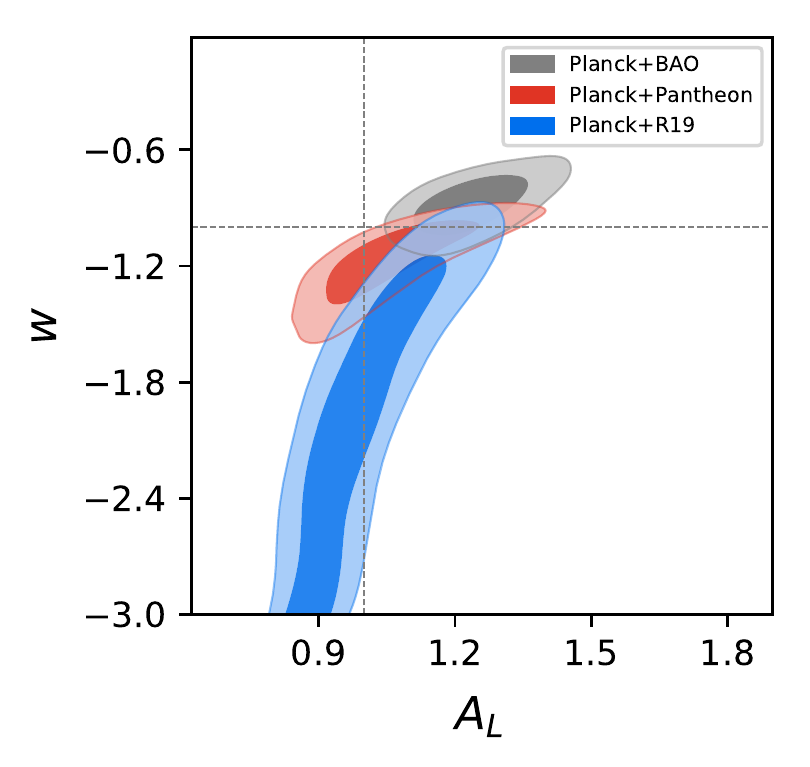}
\includegraphics[width=.38\textwidth,keepaspectratio]{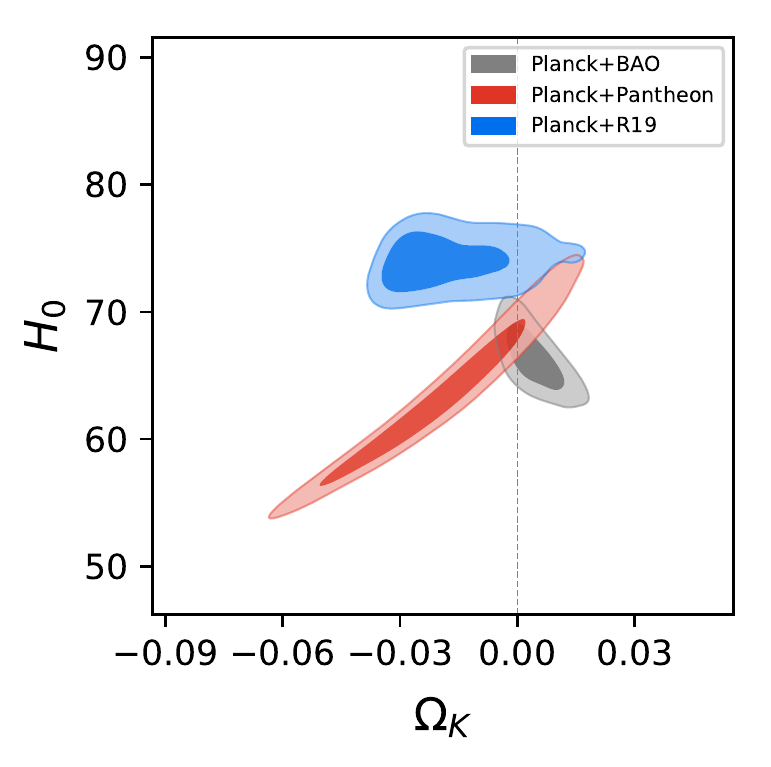}
\caption{{\bf Check for systematics}. Constraints at the $68\%$ and $95\%$ CL on the $A_{lens}$ vs $w$  (Right panel) and the $\Omega_k$ vs $H_0$ (Left panel) planes for the Planck+R19, Planck+BAO, and Planck+Pantheon datasets. A $9$ parameter model, LCDM+$w$+$\Omega_k$+$A_{lens}$, is assumed in the analysis. When $A_{lens}$ is included cosmic concordance is recovered, but Planck+BAO needs $A_{lens}>1$ at more than $99\%$ C.L., i.e. suggesting the presence of a significant systematic in Planck CMB angular spectra. \label{fig3}}
\end{figure}

Let us now consider the results for the $\Omega_k$+$w$+$A_{lens}$ case, as reported in the last three colums of Table~\ref{tab1} and in Figure~\ref{fig3}. As we can immediately see from Figure~\ref{fig3} the inclusion of the $A_{lens}$ parameter not only reconciles the Planck+Pantheon with the Planck+BAO dataset but these two cases are now in agreement with a flat LCDM model. The Planck+R19 dataset still disagrees with Planck+BAO and Planck+Pantheon. The inclusion of the $A_{lens}$ parameter, motivated by the possibility of a systematic in Planck, therefore reconciles current data with flat LCDM but does not solve the Hubble tension. It is important to stress that with the introduction of the $A_{lens}$ parameter, the BAO dataset is in much better compatibility with Planck data with an increase of just $\Delta \chi^2 \sim 6.5$ in the best-fit value. The Planck+BAO dataset requires $1.47 > A_{lens} > 1.05$ at $99\%$ CL.

\section{Conclusions}

In this {\it Letter}, we have shown that a combined analysis of the recent Planck angular power spectra with different luminosity distance measurements is in strong disagreement (at more than $99\%$ CL) with the two main expectations of the standard LCDM model, i.e., a flat universe and a cosmological constant. While the disagreement of Planck and R19 combined datasets with the expectations of the flat LCDM model has been already extensively discussed in the literature, the tension of the Planck+Pantheon result is new and clearly deserves further investigation.

The first question we have to address is whether any of the Planck+luminosity distance cosmologies, despite being incompatible with the BAO dataset, could agree with other, independent measurements.  As we can see from Table~\ref{tab1}, the constraints obtained in the case of Planck+F20, Planck+R19, and Planck+Pantheon for the remaining cosmological parameters are reasonable. For example, a value of the matter density in the range $0.25<\Omega_m<0.35$, acceptable in the case of galaxy cluster analyses, is compatible to one standard deviation with all cases. The derived age of the Universe is now around $t_0\sim 15$ Gyrs, allowing better compatibility with the ages of the oldest Population II stars \cite{vander}. The Planck+Pantheon result is fully compatible with the luminosity distances of high redshift quasars as presented in \cite{Risaliti:2018reu}.
There is also one other issue  worthy of mention. The CMB low multipole values and alignments  (\cite{Gruppuso:2017nap}) and the $2$-point  angular correlation function (\cite{Copi:2018wsv}) present possible discrepancies with the standard LCDM model. These anomalies are claimed to be significant (\cite{Schwarz:2015cma}) but disputed by the Planck collaboration (\cite{Akrami:2019bkn}), with any difference in these results depending on masking model uncertainties, among other issues. Nor does this effect depend on galactic plane orientation (see e.g. \cite{Natale:2019dqm}). However, the only generic explanation for lack of large angular scale correlation invokes a closed universe.

The second question is whether experimental systematics can explain most of the observed discrepancies with flat LCDM. The answer to this question is affirmative. If we assume that the anomaly is in the Planck data and that this systematic can be faithfully described by the $A_{lens}$ parameter, we have shown that the flat LCDM model is again in agreement with all combined analyses. It is however important to emphasize that the Planck+BAO data provides an evidence for $A_{lens}>1$ at more than $99 \%$ CL, .i.e. that cosmic concordance can be recovered only by paying the price of a significant systematic in Planck data.

What is the nature for $A_{lens}>1$ in Planck data is still matter of discussion. In this work, we use the Planck 2018 nominal (official) likelihood based on \texttt{Plik}. An alternative likelihood code exists, \texttt{CamSpec}, that gives results more compatible with the LCDM scenario, especially in its last version, as presented in \cite{Efstathiou:2019mdh}. When the \texttt{CamSpec} code is adopted, current tensions at about $99 \%$ CL could shift by one standard deviation to about $95 \%$ CL, i.e., in the realm of a possible statistical fluctuation. While the indication for a closed universe and $A_{lens}>1$ is also present in \texttt{CamSpec}, this shows that small shifts in the parameters could be expected when considering a different approach to the Planck likelihood. However, we note here that the \texttt{Plik} likelihood is the official likelihood validated by the Planck team. There is, therefore, no motivation at the moment to choose \texttt{CamSpec} over \texttt{Plik} apart  from any theoretical prejudice for LCDM. We also note that the indication for $A_{lens}>1$ is substantially increased also in case of CAMSPEC when BAO data are included.

Systematics can be undoubtedly present in luminosity distance data, and the tension between the values on the Hubble constant from F20 and R19 seems to point in this direction. A change in how systematics are considered in the Pantheon dataset could affect our results (see, e.g.  \cite{Martinelli:2019krf}). Nevertheless, Planck+BAO data, despite the tension between the two measurements and the strong indication for $A_{lens}>1$, clearly prefers a flat LCDM model. However, the BAO datapoints used in our analysis have been derived under flat LCDM and are, therefore, not strictly model-independent as are the CMB and luminosity distances data. For example, if the dark energy component is different from a cosmological constant and interacts with the dark matter, then non-linearities could behave differently from what is expected in LCDM and consequently affect the BAO result (see e.g. \cite{Anselmi:2018vjz,Heinesen:2019phg}). We are, therefore, in a situation where there is no apparent reason to trust one dataset more than another. 

The final question is whether a closed model with a phantom ($w<-1$) dark energy component is theoretically appealing. Closed inflationary models have been proposed in the literature \cite{Linde:2003hc} and a closed universe is expected in several scenarios (see e.g. \cite{Ellis:2002we}, \cite{Barrow:2010xt},\cite{Novello:2008ra}). An experimental indication for a phantom dark energy component could hint for interaction between dark matter and a $w>-1$ dark energy component (see e.g. \cite{Das:2005yj,Wang:2016lxa}). In this respect, it is interesting to note that if a closed universe increases the fine-tuning of the theory, the removal of a cosmological constant, on the other hand, reduces it. It is, therefore difficult to decide whether a phantom closed model is less or more theoretically convoluted than LCDM. 

Our conclusions are that, taken at face value, the Planck data provide a significant indication against the flat LCDM scenario, especially when combined with luminosity distance measurements. Not fitting practically half of the current cosmological data is undoubtedly a significant blow to the LCDM model. Moreover, the tensions that we have found significantly affect the ability of cosmology to test fundamental physics. For example, considering Table~\ref{tab1}, we can see that a Planck+F20 analysis indicates a neutrino mass of $\Sigma m_{\nu}\sim 0.3$ eV at the level of one standard deviation, while Planck+BAO rules this out with a $95 \%$ CL limit of $\Sigma m_{\nu}<0.194$ eV. This however also means that future laboratory measurements of a neutrino mass could play a key role in resolving current cosmological tensions.
We have also show that a possible way to save LCDM is to assume a significant systematic in the Planck data. When combined with BAO, we have found that Planck+BAO suggest a value for $A_{lens}>1$ well above the $99 \%$ CL. This also clearly means that all current cosmological constraints obtained using the Planck+BAO data should be considered with some caution.

In practice, either LCDM is ruled out, either a systematic in the Planck angular spectra data must be present.

In conclusion, our result calls for new observations and stimulates the investigation of alternative theoretical models and solutions.

\acknowledgments

EDV is supported from the European Research Council in the form of a Consolidator Grant 
with number 681431. AM thanks the University of Manchester and the Jodrell Bank Center for Astrophysics for hospitality.  AM is supported by TASP, iniziativa specifica INFN. 

\bibliography{biblio}
\bibliographystyle{aasjournal}

\end{document}